\documentclass[]{aastex631}
\usepackage{graphicx}
\usepackage{natbib}
\usepackage{aas_macros}
\usepackage{wasysym}

\shortauthors{Cordiner et al.}

\begin{document}

\title{Evidence for Surprising Heavy Nitrogen Isotopic Enrichment in Comet 46P/Wirtanen's Hydrogen Cyanide}

\correspondingauthor{M. A. Cordiner}
\email{martin.cordiner@nasa.gov}


\author[0000-0001-8233-2436]{M. A. Cordiner}
\affiliation{Astrochemistry Laboratory, NASA Goddard Space Flight Center, 8800 Greenbelt Road, Greenbelt, MD 20771, USA.}
\affiliation{Department of Physics, Catholic University of America, Washington, DC 20064, USA.}

\author{K. Darnell}
\affiliation{Astrochemistry Laboratory, NASA Goddard Space Flight Center, 8800 Greenbelt Road, Greenbelt, MD 20771, USA.}
\affiliation{Department of Physics, Catholic University of America, Washington, DC 20064, USA.}

\author[0000-0002-8130-0974]{D. Bockel{\'e}e-Morvan}
\affiliation{LESIA, Observatoire de Paris, Universit{\'e} PSL, CNRS, Sorbonne Universit{\'e}, Universit{\'e} de Paris, 5 place Jules Janssen, F-92195 Meudon, France.}

\author[0000-0002-6006-9574]{N. X. Roth}
\affiliation{Astrochemistry Laboratory, NASA Goddard Space Flight Center, 8800 Greenbelt Road, Greenbelt, MD 20771, USA.}
\affiliation{Department of Physics, Catholic University of America, Washington, DC 20064, USA.}

\author[0000-0003-2414-5370]{N. Biver}
\affiliation{LESIA, Observatoire de Paris, Universit{\'e} PSL, CNRS, Sorbonne Universit{\'e}, Universit{\'e} de Paris, 5 place Jules Janssen, F-92195 Meudon, France.}

\author[0000-0001-7694-4129]{S. N. Milam}
\affiliation{Astrochemistry Laboratory, NASA Goddard Space Flight Center, 8800 Greenbelt Road, Greenbelt, MD 20771, USA.}

\author[0000-0001-6752-5109]{S. B. Charnley}
\affiliation{Astrochemistry Laboratory, NASA Goddard Space Flight Center, 8800 Greenbelt Road, Greenbelt, MD 20771, USA.}

\author[0000-0002-1545-2136]{J. Boissier}
\affiliation{Institut de Radioastronomie Millimetrique, 300 rue de la Piscine, F-38406, Saint Martin d'Heres, France.}

\author[0000-0002-6391-4817]{B. P. Bonev}
\affiliation{Department of Physics, American University, Washington D.C., USA.}

\author[0000-0001-8642-1786]{C. Qi}
\affiliation{Harvard-Smithsonian Center for Astrophysics, 60 Garden Street, MS 42, Cambridge, MA 02138, USA.}

\author[0000-0003-2414-5370]{J. Crovisier}
\affiliation{LESIA, Observatoire de Paris, Universit{\'e} PSL, CNRS, Sorbonne Universit{\'e}, Universit{\'e} de Paris, 5 place Jules Janssen, F-92195 Meudon, France.}

\author[0000-0001-9479-9287]{A. J. Remijan}
\affiliation{National Radio Astronomy Observatory, Charlottesville, VA 22903, USA.}

\begin{abstract}
46P/Wirtanen is a Jupiter-family comet, probably originating from the Solar System's Kuiper belt, that now resides on a 5.4 year elliptical orbit. During its 2018 apparition, comet 46P passed unusually close to the Earth (within 0.08 au), presenting an outstanding opportunity for close-up observations of its inner coma. Here we present observations of HCN, H$^{13}$CN and HC$^{15}$N emission from 46P using the Atacama Compact Array (ACA). The data were analyzed using the SUBLIME non-LTE radiative transfer code to derive $^{12}$C/$^{13}$C and $^{14}$N/$^{15}$N ratios. The HCN/H$^{13}$CN ratio is found to be consistent with a lack of significant $^{13}$C fractionation, whereas the HCN/HC$^{15}$N ratio of $68\pm27$ (using our most conservative $1\sigma$ uncertainties), indicates a strong enhancement in $^{15}$N compared with the solar and terrestrial values. The observed $^{14}$N/$^{15}$N ratio is also significantly lower than the values of $\sim140$ found in previous comets, implying a strong $^{15}$N enrichment in 46P's HCN. This indicates that the nitrogen in Jupiter-family comets could reach larger isotopic enrichments than previously thought, with implications for the diversity of $^{14}$N/$^{15}$N ratios imprinted into icy bodies at the birth of the Solar System.
\end{abstract}


\section{Introduction}

Comets consist of a mixture of ice, dust and pebbles, which are thought to have accreted in the vicinity of the giant planets around 4.5 Gyr ago, and have remained relatively unaltered ever since. Measurements of their compositions therefore provide a unique tool for investigating chemical and physical processes that occurred in the protosolar accretion disk during (and prior to) the epoch of planet formation. Due to the difficulty of protoplanetary disk midplane observations using even the most powerful ground and space-based telescopes, important details regarding the chemistry of star and planet formation remain unknown \citep{obe23}. Cometary observations are uniquely useful in their ability to provide fundamental, quantitative constraints on astrochemical models for star and planet forming regions, in particular, regarding the chemistry that occurred during the earliest history of our Solar System. 

Isotopic ratios such as D/H and $^{15}$N/$^{14}$N within cometary molecules are especially sensitive to the physical conditions prevalent during the formation and accretion of cometary matter. Isotopic fractionation is the process by which different isotopes of a given atom can become concentrated in a (gas or solid phase) molecular reservoir, leading to isotopic abundance ratios that differ from the elemental ratios of the bulk reservoir.  As described in the review by \citet{nom23}, isotopic fractionation occurs in interstellar, protostellar, protoplanetary disk, and planetary environments through a broad range of gas- and solid-phase processes.

In dense interstellar clouds, strong depletion of $^{15}$N in N$_2$H$^+$ gas is commonly found relative to the local interstellar medium (ISM) \citep{biz13,red18}. On the other hand, \citet{hil18} found evidence for $^{15}$N \textit{enrichment} in HC$_{3}$N towards the L1544 prestellar core. ALMA observations of protoplanetary disks have recently revealed significant $^{15}$N enrichment in gas-phase HCN \citep{guz17,hil19}, and this can be explained as a result of isotope-selective photodissociation of N$_2$ \citep{nom23}. As shown by \citet{vis18}, self shielding of the dominant N$_2$ isotopologue leads to a region of the disk enriched in gas-phase $^{15}$N, which becomes incorporated into other gas-phase molecules, resulting in enhanced $^{15}$N/$^{14}$N ratios. When the density is high enough and the temperature is low enough, such isotopically enriched gas-phase molecules freeze out onto dust grain surfaces to form ice mantles, which are later incorporated into comets and other icy bodies.

Consistent with this picture, the observed protoplanetary disk HC$^{14}$N/HC$^{15}$N ratios of $\sim100$--200 are similar to those found in comets \citep{nom23}, which corroborates our basic understanding of the genetic relationship between protoplanetary disk and cometary compositions. Unlike protoplanetary disks, however, comets show a surprising degree of uniformity in their $^{14}$N/$^{15}$N ratios (among different molecules, and across different comets), with a weighted average value of $144\pm3$ from HCN, CN and NH$_2$ in 31 comets \citep{hil17}.  A similar nitrogen isotopic fingerprint was also found in the molecular nitrogen gas emitted by comet 67P ($^{14}$N/$^{15}$N $\sim 130\pm30$; \citealt{alt19}).

Continued studies of cometary isotopic ratios are therefore of interest, to explore the distribution of  $^{14}$N/$^{15}$N values for comparison with observations of protoplanetary disks and models for the formation of our own Solar System, with the aim of constraining the physics and chemistry of these crucial planet-forming environments.  In this article, we present new results on the HC$^{14}$N/HC$^{15}$N abundance ratio in Jupiter family comet 46P/Wirtanen, which was observed by \citet{cor23} using the Atacama Large Millimeter/submillimeter Array (ALMA) during its exceptional 2018 apparition. The unusually close Earth-comet distance allowed the detection of weak spectral lines not typically detectable in Jupiter family comets from the ground, resulting in the first map of HC$^{15}$N in a Jupiter family comet, and new insights into the possible diversity of $^{14}$N/$^{15}$N ratios among the comet population.

\section{Observations}

Observations of comet 46P/Wirtanen were conducted using ALMA during 2018 December 2-7, when the comet was around 0.1 au from Earth and 1.06--1.07~au from the Sun (the comet's perihelion date was 2018-12-12). This study focuses on the Atacama Compact Array (ACA) data, which incorporated $12\times7$~m antennas covering baselines in the range 9-50 m. The shorter baselines of the ACA compared with the main (12~m) ALMA array make it more sensitive to extended coma emission; the resulting synthesized beam size (angular resolution) was $\theta_B=4.5''\times2.8''$ at 354 GHz. Observations were conducted of the HCN ($J=4-3$), H$^{13}$CN ($J=3-2$) and HC$^{15}$N ($J=3-2$) transitions, using the Band 6 and 7 receivers. Multiple lines of the CH$_3$OH $J_K=5_K-4_K$ band were observed in the range 241--242 GHz in order to derive the coma kinetic temperature. Additional observational parameters are given in Table \ref{tab:lines}, including the spectral resolution ($\Delta{\nu}$), Geocentric distance ($\Delta$) and spectrally integrated line intensity ($\int{S_{\nu}}d{v}$, with $\pm1\sigma$ statistical errors derived from the actual noise level {inside a $\pm90$~km\,s$^{-1}$ region adjacent to each spectral line}). For CH$_3$OH, the integrated line intensity was summed over the 14 detected transitions of the $J_K=5_K-4_K$ band.

\begin{table*}
\begin{center}
\caption{Observed Spectral Line Details \label{tab:lines}}
\begin{tabular}{lccccccr}
\hline
Species & Transition & Freq.$^a$ & Obs. Date & $\theta_B$ & $\Delta{\nu}$ & $\Delta$ &$\int{S_{\nu}}d{v}$  \\      &    ($J''-J'$)      &  (GHz)&     (UT)  & ($''$) & (kHz) & (au) & (mJy\,km\,s$^{-1}$)\\
\hline
HCN       &$4-3$&354.505477&2018-12-02 02:59 & $4.5\times2.8$   & 122  & 0.12 &$2,806\pm37$\\
HC$^{15}$N&$3-2$&258.156996&2018-12-07 23:58 & $7.6\times4.1$   & 977  & 0.09 &$41\pm9$\\
H$^{13}$CN&$3-2$&259.011798&2018-12-07 23:58 & $7.6\times4.1$   & 977  & 0.09 &$32\pm10$\\
CH$_3$OH  & $5_K-4_K$&241.7--241.9&2018-12-07 23:58& $7.0\times4.2$ & 244 & 0.09& $3,472\pm68$\\
\hline
\end{tabular}
\end{center}
\parbox{0.9\textwidth}{\footnotesize 
\vspace*{1mm}
$^a$ Spectral line frequencies were obtained from the CDMS database \citep{end16}. For the HCN isotopologues, only the strongest hyperfine component is given.\\
}
\end{table*}

Data flagging, calibration and continuum subtraction were performed as described by \citet{cor23}. Imaging was performed using the CASA {\tt tclean} (H{\"o}gbom) algorithm with natural weighting and a pixel size of $0.5''$. Deconvolution was carried out within a $30''$-diameter circular mask centered on the comet, with a flux threshold of twice the RMS noise level ($\sigma$).

\section{Results}

\begin{figure}
\begin{center}
 \includegraphics[width=0.5\columnwidth]{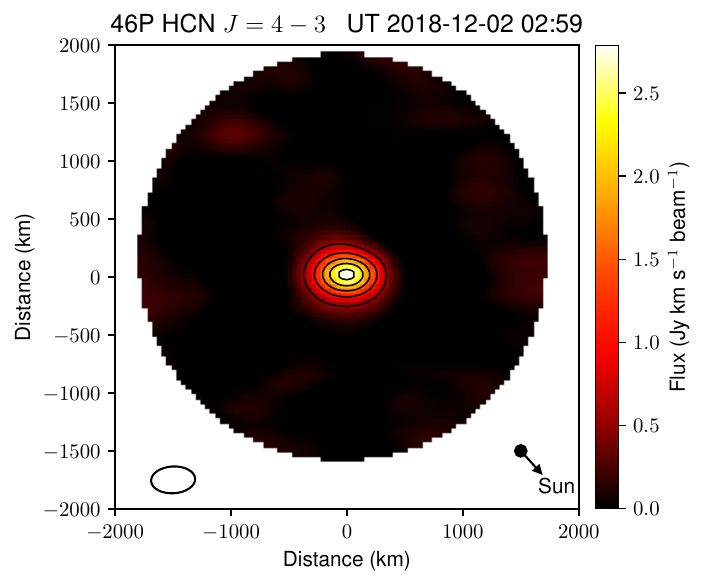}\\
 \includegraphics[width=0.5\columnwidth]{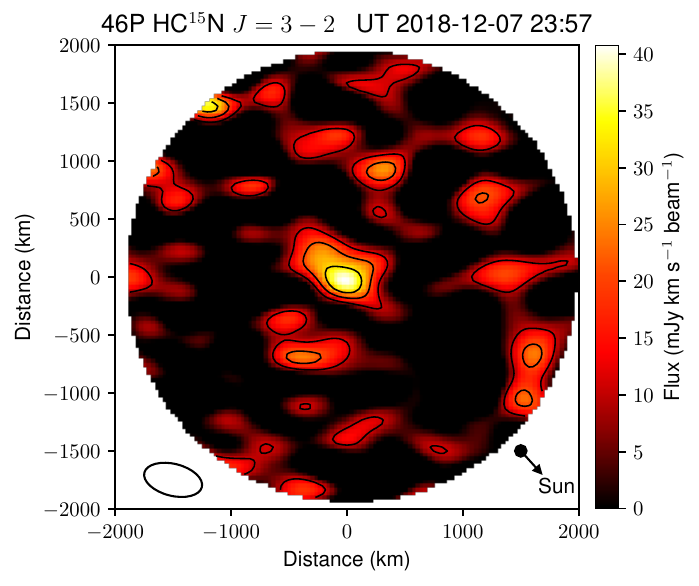}\\
 \includegraphics[width=0.5\columnwidth]{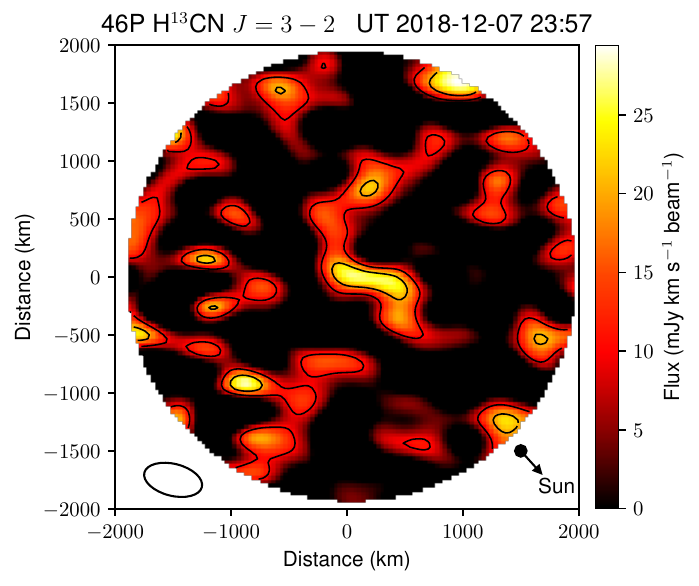} 
 \caption{ALMA ACA maps of spectrally integrated HCN, H$^{13}$CN and HC$^{15}$N emission from comet 46P/Wirtanen. Contour intervals are in units of $5\sigma$ for HCN and $1\sigma$ for H$^{13}$CN and HC$^{15}$N. Beam size (angular resolution) is shown lower left; {sky-projected comet-sun vectors are shown lower right}. The HCN map is centered on the emission peak, whereas the H$^{13}$CN and HC$^{15}$N maps are centered on the stronger (simultaneously observed) CH$_3$OH emission peak (not shown).}
   \label{fig:maps}
\end{center}
\end{figure}

Spectrally integrated flux maps for the three HCN isotopologues are shown in Fig. \ref{fig:maps}, integrated over velocity ranges $\pm1.2$~km\,s$^{-1}$ with respect to the line rest velocities. Angular distances on the sky have been converted to spatial coordinates at the distance of the comet, with the origin at the HCN peak (for HCN), and at the CH$_3$OH peak for HC$^{15}$N and H$^{13}$CN (observed simultaneously with CH$_3$OH). For the weaker (HC$^{15}$N and H$^{13}$CN) lines, the spectral integration ranges were determined based on the velocity width of the stronger (HCN) line.

\begin{figure}
\begin{center}
 \includegraphics[width=0.5\columnwidth]{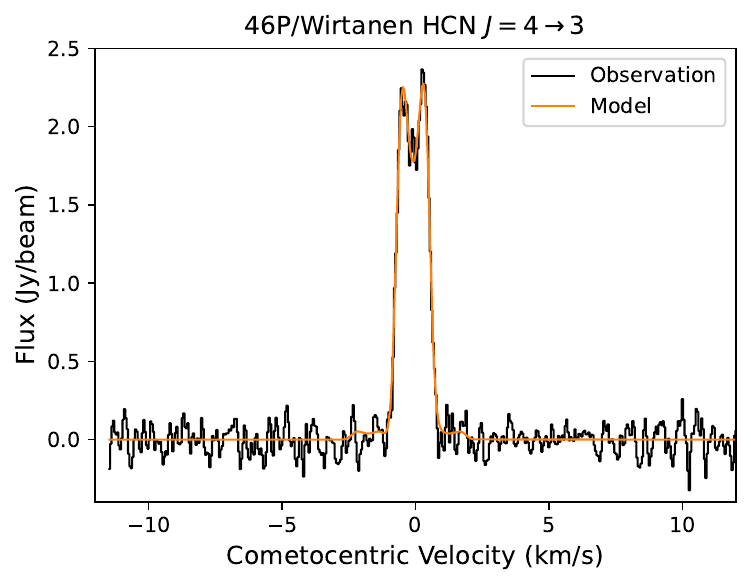}\\
 \includegraphics[width=0.5\columnwidth]{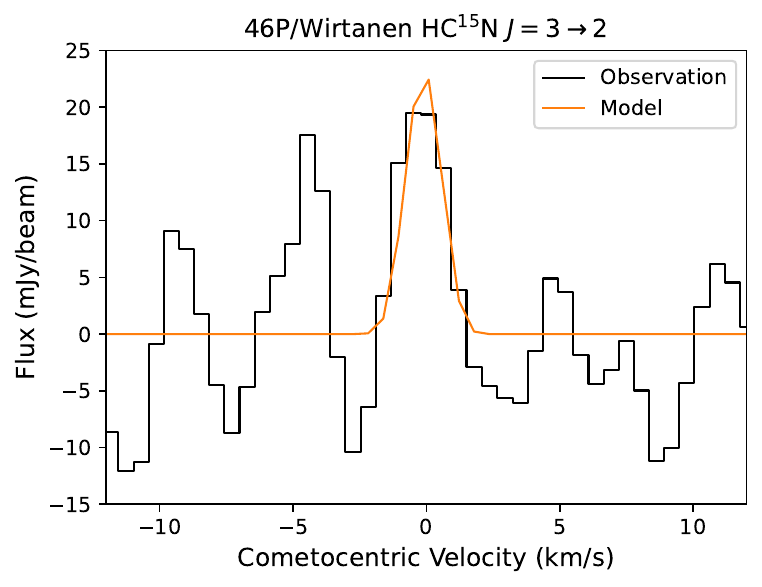} \\
 \includegraphics[width=0.5\columnwidth]{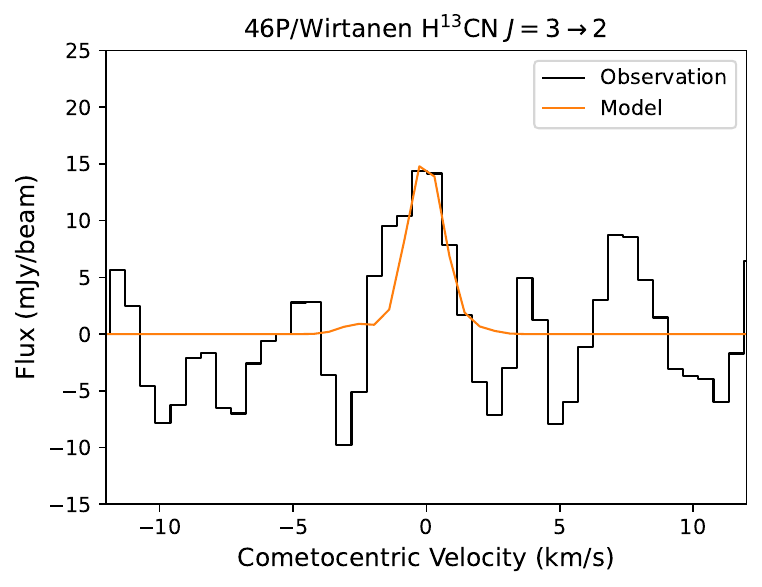} 
 \caption{ALMA ACA spectra of HCN, H$^{13}$CN and HC$^{15}$N in comet 46P/Wirtanen, extracted from the central position(s) shown in Figure \ref{fig:maps}. Best fitting SUBLIME radiative transfer models (including hyperfine structure) are overlaid in orange. The $3\sigma$ feature in the HC$^{15}$N spectrum around $-5$~km\,s$^{-1}$ is interpreted as an unusually strong ($>3\sigma$) noise spike.}
   \label{fig:spectra}
\end{center}
\end{figure}

\begin{figure*}
\begin{center}
 \includegraphics[width=\textwidth]{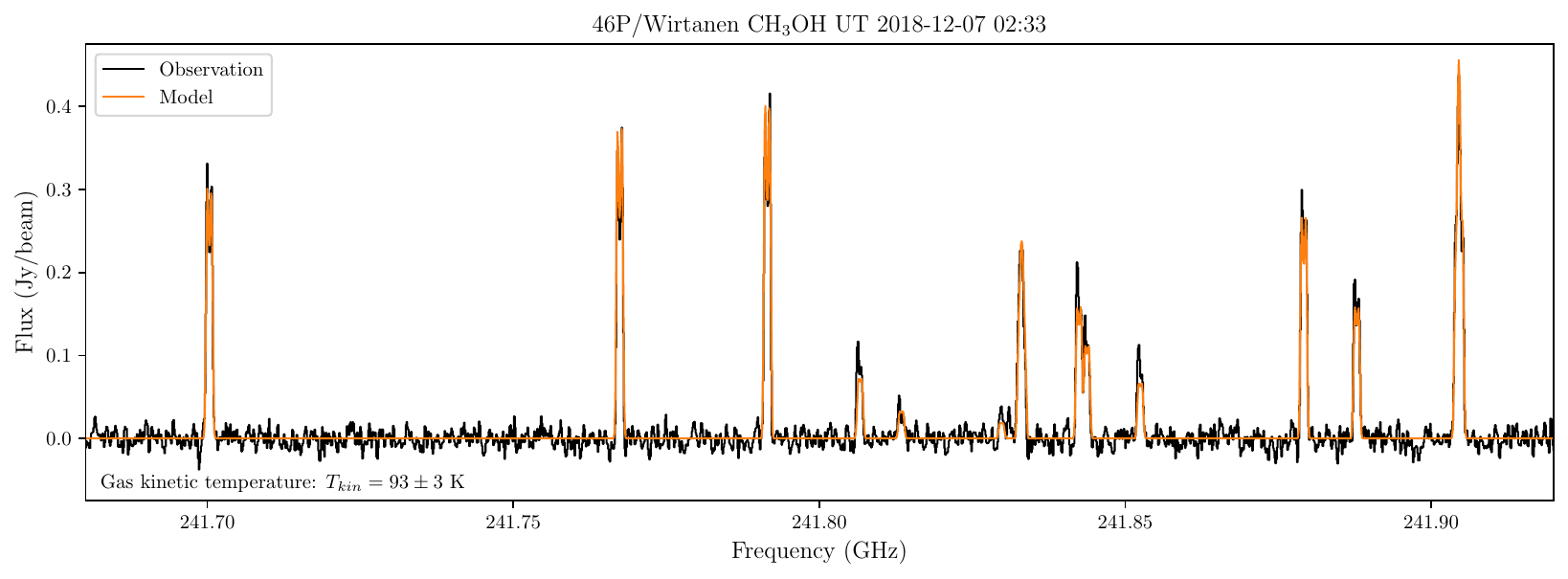}
 \caption{ALMA ACA spectrum of CH$_3$OH observed in comet 46P/Wirtanen (extracted from the CH$_3$OH spectrally integrated brightness peak). Best fitting SUBLIME radiative transfer model is overlaid in orange, demonstrating a gas kinetic temperature of $T_{kin}=93\pm3$~K.}
   \label{fig:ch3oh}
\end{center}
\end{figure*}

Spectra were extracted from the (0,0) position in each map (shown in Fig. \ref{fig:spectra}). {Based on the spectrally integrated line fluxes}, HCN is detected at a high significance ($76\sigma$), while HC$^{15}$N and H$^{13}$CN are detected at $4.6\sigma$ and $3.2\sigma$, respectively. The observed CH$_3$OH spectrum is shown in Fig. \ref{fig:ch3oh}. The spectral line profiles are well resolved for HCN and CH$_3$OH, showing a characteristic double-peaked sub-structure due to Doppler motion of the quasi-isotropically expanding coma along the line of sight, whereas no sub-structure is expected for the lower-resolution H$^{13}$CN and HC$^{15}$N observations. 

The raw statistical significance of our HC$^{15}$N and H$^{13}$CN detections is less than optimal, but the evidence for both molecules is strengthened by the properties of the spectral line profiles, which match (within the noise) the rest velocity and FWHM of the (high-significance) HCN line. In the case of HC$^{15}$N, the four spectral channels that make up the line peak are all at a significance of greater than $3\sigma$ (where $\sigma=4.7$~mJy). Furthermore, the HC$^{15}$N emission peak coincides with the known location of the comet's nucleus, as derived from the peak position of the spectrally integrated CH$_3$OH data. As a general rule of thumb in radio interferometry, a detection can be considered real if appears above the $3\sigma$ level for a source with a known location (as is the case here), or above $5\sigma$ if the location is unknown; both HC$^{15}$N and H$^{13}$CN therefore fulfill the detection criteria.

\begin{table*}
\begin{center}
\caption{Production rates and abundances for H$_2$O and the observed 46P HCN isotopologues \label{tab:results}}
\begin{tabular}{lcccc}
\hline
Species & Obs. Date & $Q({\rm H_2O})$ (s$^{-1}$) & $Q({\rm Species})$ (s$^{-1}$) & Abundance \\
\hline
HCN       &2018-12-02 &  $5.67\times10^{27}$  &  $(5.20\pm0.09)\times10^{24}$   &$(9.17\pm0.16)\times10^{-4}$\\
HC$^{15}$N&2018-12-07 &  $7.56\times10^{27}$  &  $(1.03\pm0.25)\times10^{23}$   &$(1.36\pm0.33)\times10^{-5}$\\
H$^{13}$CN&2018-12-07 &  $7.56\times10^{27}$  &  $(7.70\pm2.49)\times10^{22}$   &$(1.03\pm0.32)\times10^{-5}$\\
\hline
\end{tabular}
\end{center}
\end{table*}

Spectral modeling was performed using the spherically symmetric (1D) version of SUBLIME: a time-dependent, non-LTE radiative transfer code for cometary comae \citep{cor22}, and the 1D version of the model was found to provide a sufficiently good fit to these ACA data. In the molecular excitation calculation, collision rates between CH$_3$OH and H$_2$O were assumed to be the same as CH$_3$OH with H$_2$ \citep{rab10}. HCN--H$_2$O collision rates are from \citet{dub19}, and were assumed to apply equally for all three HCN isotopologues. Rovibrational pumping due to the solar radiation field was calculated using molecular data for HCN and CH$_3$OH from the HITRAN and Planetary Spectrum Generator databases \citep[see][for details]{cor23,vil18}. Hyperfine structure was included in the model spectra for all three HCN isotopologues, assuming equilibrium line-strength ratios among the hyperfine components in a given $J$ state. 

H$_2$O production rates for comet 46P ($Q({\rm H_2O})$) were obtained for our observation times from cubic spline interpolation of the SOHO Ly-$\alpha$-derived measurements by \citet{com20}, and are given in Table \ref{tab:results}. {These $Q({\rm H_2O})$ values are consistent with the average water production rate of $8\times10^{27}$~s$^{-1}$ measured by \citet{lis19} using the SOFIA telescope a few days closer to perihelion (between December 14--20), and with the average value of $7\times10^{27}$~s$^{-1}$ obtained using IRTF between December 6--21 (Khan et al. 2023).  The fact that the values from three different observatories (spanning ultraviolet to far-infrared wavelengths) are consistent despite the large differences in their beam sizes, implies that the SOHO water production rates should be applicable at the spatial resolution of our ACA data.}

Prior to spectral modeling, the measured ACA spectra were corrected for interferometric flux loss factors of 0.79 for HCN and 0.73 for HC$^{15}$N and H$^{13}$CN. These were derived based on initial best-fit SUBLIME model image cubes, which were spectrally integrated then processed using CASA {\tt simobserve} according to the particular sky position and observation time of each line, then subject to the same cleaning and deconvolution as the observations. The purpose of this loss factor is to account for the fact that the the ACA images are missing a portion of the flux from the extended coma because the interferometer is only sensitive to structures less than $\lambda/D$ in spatial extent, where $\lambda$ is the observed wavelength and $D$ is the minimum antenna separation (9~m in this case).

A beam-averaged gas kinetic temperature of $T_{kin}=93\pm3$~K was derived using a least-squares fit to the CH$_3$OH spectrum (Fig. \ref{fig:ch3oh}). By fitting the HCN $J=4-3$ line profile, a coma outflow velocity of $0.53\pm0.01$ km\,s$^{-1}$, and Doppler shift of $-0.097\pm0.006$ km\,s$^{-1}$ was derived; these values were used in subsequent fits to the spectrally less well resolved (and lower signal-to-noise) HCN isotopologue lines. Best fitting production rates and abundances (relative to H$_2$O) for the three HCN isotopologues are given in Table \ref{tab:results}. Formal ($1\sigma$) uncertainties were derived from the diagonal elements of the covariance matrix of the least-squares fit. To account for the correlation between adjacent spectral channels introduced by the ACA correlator, the RMS noise measurement on each spectrum was multiplied by a factor of 1.29, following \citet{nix20}.

{The maximum optical depth of our radiative transfer model at the HCN central peak is 0.2, but since the coma opacity falls rapidly with nucleocentric distance, the mean optical depth inside the ACA beam is only 0.04. Hence, the HCN $J=4-3$ line is largely optically thin, as demonstrated by the ratio of the main HCN line peak with respect to the hyperfine satellites, which are too weak to be clearly detected, consistent with the optically thin limit (see Figure \ref{fig:spectra}). Considering the cometary HCN data are typically well reproduced by a spherically-symmetric coma model (see also \citealt{cor14,cor19,rot21,cor23}), it is therefore unlikely that the presence of spurious, high-opacity HCN clumps or jets would significantly impact our results.}

\section{Discussion}

46P/Wirtanen is only the second Jupiter-family comet to-date in which the minor ($^{15}$N and $^{13}$C) isotopologues of HCN have been detected. The first was 17P/Holmes, which had HCN/HC$^{15}$N = $139\pm26$ (observed using the IRAM 30-m telescope during the comet's major outburst in October 2007; \citealt{boc08}).  Accounting for purely statistical errors, the HCN/HC$^{15}$N production rate ratio in comet 46P is $67\pm16$, and the HCN/H$^{13}$CN ratio is $90\pm28$. Within the uncertainties, the $^{12}$C/$^{13}$C ratio is consistent with previous cometary observations \citep{boc15,cor19}, whereas the $^{14}$N/$^{15}$N ratio is surprisingly enriched in the minor ($^{15}$N) isotope --- the (error-weighted) average of prior HCN/HC$^{15}$N measurements in four previous comets is $146\pm11$ \citep{boc08,biv16}. A tentative ($3\sigma$) detection of HC$^{15}$N was also obtained in comet 46P by \citet{biv21} using the IRAM 30-m telescope between 2018-12-12 and 2018-12-18, leading to an HCN/HC$^{15}$N ratio of $77\pm26$. The combination of these IRAM and ALMA results add credibility to the conclusion that comet 46P's HCN was surprisingly enhanced in $^{15}$N.

Because HCN and its minor isotopologues were observed in comet 46P on different dates (almost 6 days apart), variations in the HCN/H$_2$O production rate ratio, as well as uncertainties in $Q({\rm H_2O})$ should be incorporated into our isotopic ratio uncertainties.  Based on six infrared spectroscopic measurements of $Q({\rm HCN})$ and $Q({\rm H_2O})$ in comet 46P between December 6--21 \citep{bon21,kha23}, the HCN abundance relative to H$_2$O remained apparently constant, with a standard deviation on the $Q({\rm HCN})$/$Q({\rm H_2O})$ ratio of only 0.00013 (corresponding to 6\% of the HCN/H$_2$O value), and the error-weighted mean was $Q({\rm HCN})$/$Q({\rm H_2O})$ = $0.0020\pm0.0001$. It is therefore reasonable to assume that variations in the comet's HCN/H$_2$O ratio contribute a negligible source of uncertainty to our result. Errors on $Q({\rm H_2O})$ may be more significant, however, considering the scatter in \citet{com20}'s measurements as a function of time, which amount to an RMS of $7.6\times10^{26}$~s$^{-1}$ ($\sim10$\%) with respect to the best-fitting linear trend in $Q({\rm H_2O})$ between 2018-11-28 and 2018-12-11. Adding this fractional uncertainty in quadrature with the statistical uncertainty on $Q({\rm HC^{15}N})/Q({\rm H_2O})$, combined with a further 10\% uncertainty on $Q({\rm HCN})$ and $Q({\rm HC^{15}N})$ to account for possible inaccuracies in the absolute ACA flux scale, gives a value of HCN/HC$^{15}$N = $67\pm20$.

The possibility of more extreme temporal variability in $Q({\rm H_2O})$ (or greater variability in HCN/H$_2$O) as a function of time cannot be ruled out however. In that case, we combine the error on the observed H$^{13}$CN abundance ($(1.03\pm0.32)\times10^{-5}$) with the expected HCN/H$^{13}$CN ratio ($\approx90$; based on measurements of $^{12}$C/$^{13}$C ratios in a diverse range of comets and other solar system bodies; \citealt{nom23}) to obtain a more conservative HCN abundance (and associated uncertainty) of $(9.27\pm2.88)\times10^{-3}$ at the time of our HC$^{15}$N observation. This leads to a more conservative HCN/HC$^{15}$N ratio (and uncertainty) of $68\pm27$. {For comparison with meteoritic measurements, whereby the isotopic ratio is typically expressed as a fractional enhancement of the minor isotope, $\delta^{15}$N, with respect to the terrestrial standard ratio ($^{14}$N/$^{15}$N)$_{Earth}$ \citep{nie50}, we calculate $\delta^{15}$N = ($^{14}$N/$^{15}$N)/($^{14}$N/$^{15}$N)$_{Earth} - 1$ = ($3015\pm1200$)\permil\ (per mil).}

\begin{figure}
\begin{center}
 \includegraphics[width=0.6\columnwidth]{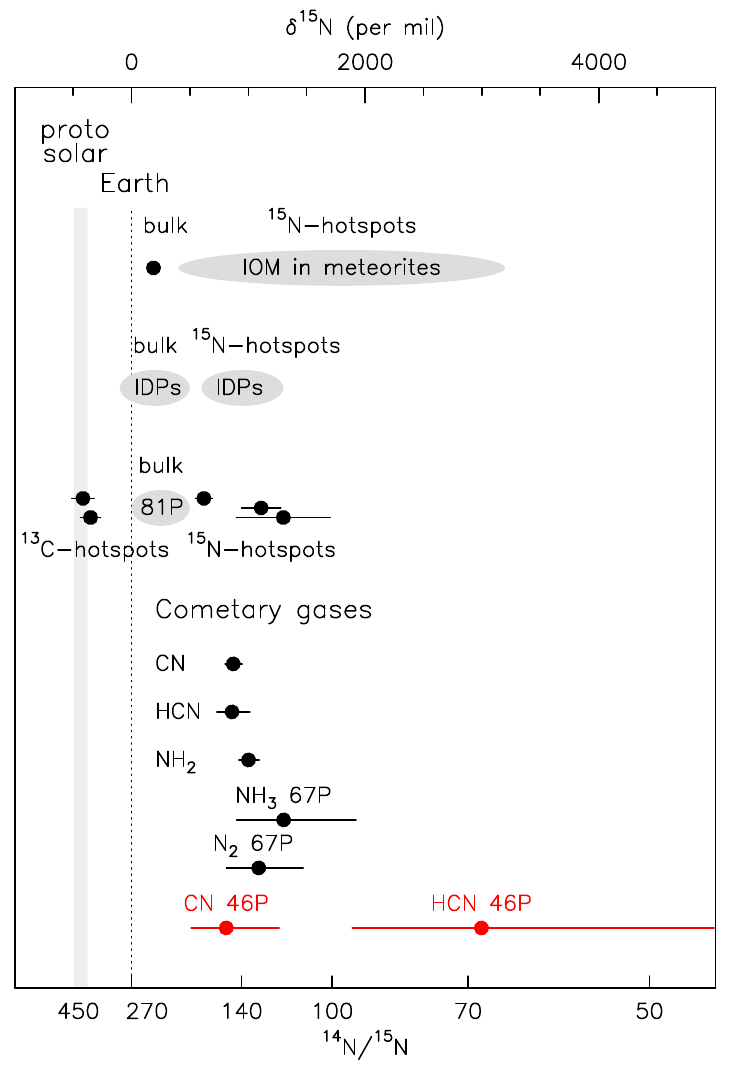}
 \caption{Nitrogen isotope measurements in primitive Solar System materials, expressed as $^{14}$N/$^{15}$N {(lower $x$ axis, with smaller values indicating greater $^{15}$N enrichment towards the right); the fractional isotopic enrichment $\delta^{15}$N is shown on the upper $x$ axis}. This Figure was adapted from \citep{boc15}. The meteoritic bulk value is from CR chondrite insoluble organic matter (IOM) measurements \citep{nom23}, while the $^{15}$N hotspots are regions that present strong isotopic enrichments relative to the surrounding meteoritic material. Grey ellipses represent the range of values from multiple laboratory sample analysis measurements. The CN, HCN and NH$_2$ measurements in cometary gases are the averages from \citet{man09}, \citet{shi16}, \citet{boc08} and \citet{biv16}. 67P values are from \citet{alt19}, and the 46P CN value is from \citet{mou23}.}
   \label{fig:nfrac}
\end{center}
\end{figure}

In the context of prior $^{14}$N/$^{15}$N ratios measured throughout the Solar System and beyond, our value in comet 46P is statistically unusual. Cometary $^{14}$N/$^{15}$N ratios are known to be systematically lower than those found in the terrestrial and giant planets, as well as the Sun, but the distribution of cometary values measured to-date appears surprisingly uniform. The weighted average $^{14}$N/$^{15}$N ratio for HCN, CN and NH$_2$ in a sample of 31 comets is $144\pm3$ \citep{hil17}, which is significantly enriched in $^{15}$N compared to the bulk terrestrial and Solar values of 273 and 459, respectively \citep{nie50,mar11}. Indeed, within the error bars, all prior cometary $^{14}$N/$^{15}$N measurements (within various molecules) may be consistent with a value $\approx140$, including the in-situ mass spectrometry measurements of the 67P coma made by the Rosetta spacecraft (see Fig. \ref{fig:nfrac}). Our new $^{14}$N/$^{15}$N value in comet 46P is $2.8\sigma$ less than the weighted average of 144, and therefore represents an unexpected outlier with respect to the overall comet population, as well as to other (bulk) Solar System bodies.  

Meteoritic organics typically have $^{14}$N/$^{15}$N ratios in the range 200--270 (somewhat enriched relative to the Earth; see Fig. \ref{fig:nfrac}). Carbonaceous chondrite meteorites also contain small, micron-sized isotopic ``hot spots'', which exhibit strong $^{15}$N enrichment (with $^{14}$N/$^{15}$N values as low as $65\pm14$; \citealt{bus06}), which is similar to our 46P value. Considering the anomalous nature of our value compared with the numerous previous cometary $^{14}$N/$^{15}$N measurements, it is interesting to speculate that comets (or comet 46P in particular) could also contain isotopically heterogeneous material. Our HC$^{15}$N observation occurred over a time period of 1 h (between UT 2018-12-07 23:58 and 2018-12-08 00:59) with a beam size $\approx3.5''$ (probing radial distances $\sim114$~km from the nucleus). For an outflow velocity of 0.53~km\,s$^{-1}$ and volatile mass loss rate of 320~kg\,s$^{-1}$ (based on $Q({\rm H_2O})$; Table \ref{tab:results}, and a typical CO$_2$/H$_2$O ratio of 17\%; \citealt{oot12}), the mass of volatiles within the ACA beam was $\sim69,000$~kg. This is indeed small compared with the total mass of the comet ($\sim4\times10^{14}$~kg, adopting a mean radius of 560~m and density of 0.6 g\,cm$^{-3}$), so it is plausible that our ACA measurement is representative of a spatially isolated $^{15}$N enhancement confined to a relatively small part of the comet's nucleus. This idea is supported by the fact that \citet{mou23} measured a nominal CN/C$^{15}$N ratio of $150\pm30$ using VLT ultraviolet spectroscopy only one day later (on 2018-12-09 at UT 00:41), which could be more indicative of the comet's bulk $^{14}$N/$^{15}$N value. However, it should be noted that CN in cometary comae often has an additional source that cannot be explained by HCN photolysis alone \citep{fra05,cot08}, so its nitrogen isotopic ratio need not necessarily be the same as HCN. As shown by Fig. \ref{fig:nfrac}, there is no prior evidence for differing CN/C$^{15}$N and HCN/HC$^{15}$N ratios in comets, but this possibility should be investigated further, and could help shed light on the extent of the chemical link between CN and HCN in comets. 

Given the 9.1 h rotation period of the 46P nucleus \citep{far21}, the time difference of $\approx24$~h between our ACA HCN observations and the VLT CN observations of \citet{mou23} (corresponding to 2.6 nucleus rotations) also allows for the possibility that different regions of the nucleus have different nitrile $^{14}$N/$^{15}$N ratios, in the case of outgassing dominated by a solar-facing jet (as was deduced for this comet by \citealt{cor23}).

$^{15}$N enrichment in cometary nitriles and meteoritic organics probably occurred as a consequence of isotope-selective photodissociation of N$_2$ in the protosolar nebula \citep{vis18,lee21}. Although such fractionation appears to have operated to produce relatively uniform $^{14}$N/$^{15}$N ratios (in the range $140\sim170$) within the bulk of the icy materials found in the comets and moons of the outer Solar System \citep{fur15}, the full range of $^{14}$N/$^{15}$N values present in the planet forming regions of our protosolar disk remains unknown. The surprisingly low HCN/HC$^{15}$N ratio in comet 46P therefore presents a new challenge for our understanding of the physical and chemical processes that occurred during Solar System formation.

Although the $^{12}$C/$^{13}$C values measured in molecules in interstellar clouds and star-forming regions exhibit some genuine diversity, the $^{12}$C/$^{13}$C ratios measured throughout the Solar System are relatively uniform \citep{nom23}.  In contrast to HC$^{15}$N, our HCN/H$^{13}$CN ratio in 46P is consistent with the values of $^{12}$C/$^{13}$C $\sim90$ found in meteorites, terrestrial and giant planets, and their icy satellites, as well as other comets. Consequently, we confirm the findings of \citet{cor19} suggesting a lack of strong carbon isotopic fractionation in cometary HCN during the formation of the Solar System.

\section{Conclusion}

We observed rotational emission lines from HCN and its two minor isotopologues HC$^{15}$N and H$^{13}$CN in the coma of comet 46P during its exceptionally close December 2018 apparition, using the Atacama Compact Array. The HC$^{15}$N ($J=3-2$) line was surprisingly strong compared with the H$^{13}$CN ($J=3-2$) line, allowing HC$^{15}$N to be mapped for the first time in a comet. The spectral line data were subject to non-LTE radiative transfer modeling, from which we derived a HCN/HC$^{15}$N production rate ratio of $67\pm17$ (or a more conservative value of $68\pm27$, based on the observed H$^{13}$CN and HC$^{15}$N abundances, with an assumed HCN/H$^{13}$CN ratio of 90). These values are significantly lower than previously measured in any N-bearing molecules in comets or in any other large Solar System bodies. Within the Solar System, comet 46P's HCN was therefore surprisingly enriched in $^{15}$N, reaching a value similar to those found in the most $^{15}$N-enriched ``hotspots'' in spatially isolated meteoritic samples. This result implies that cometary $^{14}$N/$^{15}$N ratios could be more diverse, and potentially $^{15}$N rich, than previously thought. More studies of $^{14}$N/$^{15}$N in comets are warranted to better understand the diversity of isotopic fractionation processes taking place in protoplanetary disks.

\begin{acknowledgments}
This work was supported by the National Science Foundation under grant Nos. AST-2009253, AST-2009398, and by NASA’s Planetary Science Division Internal Scientist Funding Program through the Fundamental Laboratory Research work package (FLaRe). This work makes use of ALMA data set ADS/JAO.ALMA\#2018.1.01114.S. ALMA is a partnership of ESO (representing its member states), NSF (USA) and NINS (Japan), together with NRC (Canada), NSTC and ASIAA (Taiwan), and KASI (Republic of Korea), in cooperation with the Republic of Chile. The Joint ALMA Observatory is operated by ESO, AUI/NRAO and NAOJ. The National Radio Astronomy Observatory is a facility of the National Science Foundation operated under cooperative agreement by Associated Universities, Inc.
\end{acknowledgments}

\bibliography{refs}{}
\bibliographystyle{aasjournal}

\end{document}